\documentclass[11pt,final,twocolumn,twoside]{IEEEtran}
\usepackage{amsmath}
\usepackage[margin=0.75in,headheight=0.45in]{geometry}
\usepackage[pdftex]{epsfig}
\usepackage{amsfonts}
\usepackage{amssymb}
\usepackage{fancyhdr}
\include{graphicsx}

\newcommand\cu{\ensuremath{C_u}}
\newcommand\tu{\ensuremath{T_u}}
\newcommand\dtu{\ensuremath{\frac{d\tu}{dt}}}

\newcommand\cd{\ensuremath{C_d}}
\newcommand\td{\ensuremath{T_d}}

\newcommand\dtd{\ensuremath{\frac{d\td}{dt}}}

\newcommand\calf{{\cal F}}

\begin{document}
\title{\vspace{0.2in}\sc Dependence of Inferred Climate Sensitivity on
  the Discrepancy Model}
\author{Balu Nadiga$^{1}$\thanks{\hspace{-.18cm}Corresponding author: B.T. Nadiga,
    balu@lanl.gov$\quad\quad\quad\quad\quad$ $^1$Los Alamos National
    Lab, Los Alamos, NM 87544}, Nathan Urban$^1$
\vspace{-0.75cm}
}

\maketitle
\thispagestyle{fancy}

\begin{abstract}
  We consider the effect of different temporal error structures on the
  inference of equilibrium climate sensitivity\footnote{ECS is defined
    as the realized equilibrium surface warming---globally-averaged
    surface air temperature---for a doubling of CO$_2$}(ECS), in the context
  of an energy balance model (EBM) that is commonly employed in
  analyzing earth system models (ESM) and observations. We consider
  error structures ranging from uncorrelated (IID normal) to AR(1) to
  Gaussian correlation (Gaussian Process GP) to analyze the abrupt
  4xCO$_2$ CMIP5 experiment in twenty-one different ESMs. For seven of
  the ESMs, the posterior distribution of ECS is seen to depend rather
  weakly on the discrepancy model used suggesting that the
  discrepancies were largely uncorrelated. However, large differences
  for four, and moderate differences for the rest of the ESMs,
  leads us to suggest that AR(1) is an appropriate discrepancy
  correlation structure to use in situations such as the one
  considered in this article.

  Other significant findings include: (a) When estimates of ECS (mode)
  were differrent, estimates using IID were higher (b) For four of the
  ESMs, uncertainty in the inference of ECS was higher with the IID
  discrepancy structure than with the other correlated structures, and
  (c) Uncertainty in the estimation of GP parameters were much higher
  than with the estimation of IID or AR(1) parameters, possibly due to
  identifiability issues.  They need to be investigated further.

\end{abstract}
\begin{figure*}[]
\centering
\includegraphics[width=\textwidth]{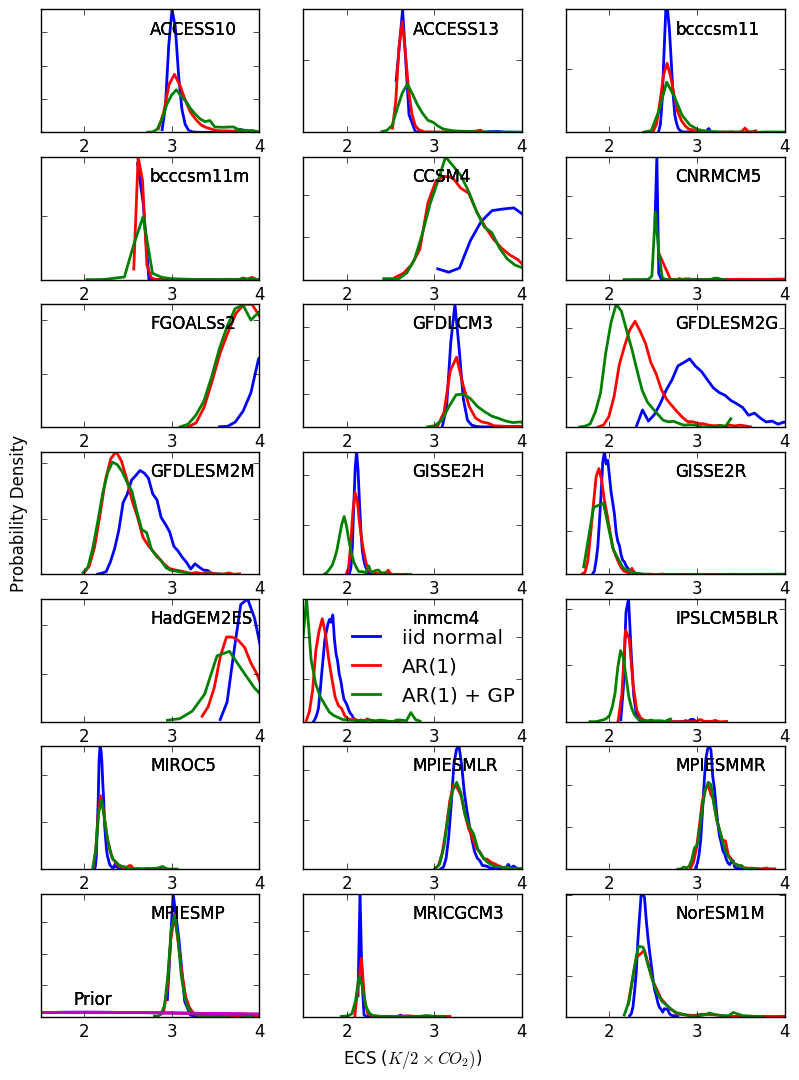}
\caption{Inferred ECS for the 21 models considered using CMIP5 experiment abrupt4xCO$_2$.}
\end{figure*}

\section{Introduction}
On the one hand, Earth System Models (ESMs) that comprise of
atmosphere-ocean general circulation models (AOGCMs) coupled to other
earth system components such as ice sheets, land surface, terrestrial
biosphere, and glaciers are central to developing our understanding of
of the workings of the climate system, and are proving to be the most
comprehensive tool available to study climate change and develop
climate projections \cite{flato2013evaluation}.  Concomitantly, simple
concepts such as climate sensitivities---metrics used to characterise
the response of the global climate system to a given forcing---are
central not only to climate modeling, but also to discussions of the
ongoing global warming (e.g., see \cite{mitchell1990equilibrium,
roe2007climate}.  Nevertheless, the immense computational
infrastructure required and the cost incurred in running ESMs
precludes the direct evaluation of such metrics.

Simple climate models (SCMs) on the other hand consider only integral
balances of important quantities such as mass and/or energy, are
computationally cheap and can be used in myriad differnt ways (unlike
ESMs that are typically run only in the forward mode). It is for these
reasons that the use of SCMs to estimate climate sensitivities, both in
the context of ESMs and actual observations, is now well established
(e.g., see \cite{hoffert1980role, held2010probing} and others).

\section{Methodology and Results}
A particular form of an SCM that has been popular in summarizing
integral thermal properties of AOGCMs and/or ESMs is the anomaly-based
upwelling-diffusion (UD) energy-balance model (EBM)
\cite{hoffert1980role}.  To briefly describe such a model, consider a
horizontally-integrated model of the climate system that is
partitioned into two active layers in the vertical. An upper (surface)
layer that that comprises the oceanic mixed layer, atmosphere and land
surface and and a bottom layer that comprises the ocean beneath the
mixed layer.  Evolution of the upper surface heat content anomaly per
unit area is given by
\begin{align}
\cu\dtu &= \calf - \lambda\tu - \gamma\left(\tu-\td\right) 
\label{eq-upper}
\end{align}
where $\lambda$ represents. Exchange of heat between the surface layer
and the  ocean beneath is parameterized by the difference in
temperature between the two layers.
Similarly, evolution of the subsurface  ocean heat content anomaly (again
per unit area ) is given by
\begin{align}
\cd\dtd &= - \gamma\left(\tu-\td\right).
\label{eq-lower}
\end{align}

Such two layer models have been used extensively to obtain point
estimates of ECS of AOGCMs/ESMs (e.g., see \cite{held2010probing,
  geoffroy2013transient}, and others). 
However, and to the best of our knowledge, the dependence of estimates
of ECS on the assumed temporal structure of the discrepancy between
ESM representation of the surface air temperature (SAT) and the above
EBM's (Eqs.~1 \& 2) representation of it has not been investigated:
\begin{align}
T_u^{ESM}(t) = T_u^{EBM}(t) + \epsilon_t
\end{align}
The equation above arises from the fact that the anomaly-based EBM
considered has no representation of climate variability, unlike the
more comprehensive ESM that it is used to analyze.
We consider three correlation structures for $\epsilon_t$:
\begin{enumerate}
\item IID: $\Sigma (t-s) = \sigma^2\delta(t-s)$ 
\item AR(1): $\Sigma (t-s) =   \sigma^2 \rho^{|t-s|}$
\item AR(1)+GP: $\Sigma (t-s) =   \sigma^2
  \exp(-\frac{(t-s)^2}{\lambda^2})$
\end{enumerate}
where structure AR(1)+GP uses the sum of the covariances indicated in items
2 and 3 above

In the context of globally-averaged SAT (of which sea surface
temperature or SST is a large
component), we know, e.g., following the work of
\cite{hasselmann1976stochastic}, that the mixed layer integrates
(high-frequency) weather noise. Thus, SST (and therefore SAT) is
expected to be auto-correlated in time, although the correlations
themselves are highly spatiotemporally variable (e.g, winter SST
anomalies are more persistent than summer SST anomalies, the tropical
Pacific may display larger persistence than the tropical Atlantic,
etc...). However, such correlations rarely exceed about six months and
we are considering annual-averaged SAT. Physically, correlations on
the interannual time scales are related to internal climate dynamics
(phenomena such as the re-emergence of the winter mixed layer,
delay-oscillations and others). While a casual inspection of some
actual climate timeseries may suggest the unlikeliness of IID
variability, it is not the case for the globally-averaged SAT
timeseries in the abrupt4xCO$_2$ CMIP5 experiment that we analyze, and
as we will see later. However, it should also be noted that if the
discrepancies are actually correlated, then an inference of EBM
parameters using the IID discrepency structure will result in estimates of
uncertainty that are smaller than actual, and again as we will see
later.

Figure 1 shows the posterior distribution of ECS with the three error
structures (indicated in the legend) for the 21 ESMs. The prior is shown
in the bottom-left panel. In this figure it is seen that
\begin{itemize}
\item For a substantial number of the ESMs, the three error structures
  lead to similar estimates of ECS (CNRMCM5, MIROC5, MPIESML/MR,
  MPIESMP, MRICGCM3, NorESM1M)
\item When the estimates of ECS are differrent, estimates using IID tend to
  be higher (CCSM4, FGOALSs2, GFDLESM2G/M, HadGEM2ES, inmcm4)
\item Differences in ECS estimates from that between AR(1) and AR(1)+GP
  tend to be smaller than that between either and IID
\item In a majority of the ESMs considered, IID leads to smaller
  estimates of uncertainty in ECS suggesting that the discrepancies in
  those models are temporally correlated. The exceptions (CCSM4,
  GOALSs2, GFDLESM2G/M), are therfore surprising and need to be
  investigated further.
\end{itemize}
We also note that there was far more uncertainty in the estimation of
GP parameters as opposed to estimateion of either IID and AR(1)
parameters.  This is likely due not only to the shortness of the ESM
runs considered (150 years) from the point of low-frequency
variability that the GP component was intended to capture, but may
involve issues of identifiability and needs to be investigated
further. However, when such problems occur, the parameters involved
act more as nuisance parameters and do not prevent reasonable
inference of ECS and other EBM parameters.

\section{Discussion}
Simple climate models play a valuable role in helping interpret both
observations and the responses of comprehensive ESMs. As such, we
used a simple and popular EBM in a Bayesian framework to interpret the
abrupt4xCO$_2$ CMIP5 experiment in 21 ESMs, in terms of their SAT
response. We used three different statistical models to represent the
discrepancy in the SAT response of the ESMs and SCMs. This discrepancy
is largely due to natural variability---an aspect of climate that
represented in the ESMs, but not in the SCMs. For seven of these
models, the posterior distribution of ECS depended only very weakly on
the discrepancy model used suggesting that the discrepancies were
largely uncorrelated. For four of the models, the differences  were large
and for the rest of the models, the differences were
moderate. Significant differences in a majority of the models,
therefore, indicate  the existence of temporal correlations in the
discrepancies and the imprtance of accounting for them in a Bayesian
inference framework.

Next, the differences in estimated ECSs were much smaller for
inferences using AR(1) and AR(1)+GP as compared to differences between
inferences using either of these models and IID. This coupled with the
fact that the uncertainty in the estimation of GP parameters was much
larger than that in the estimation of AR or IID parameters, leads us
to conclude that AR(1) is a good choice\footnote{Additionally, the
  existence of an analytic inverse for the covariance of an AR(1)
  process makes it faster to compute with as compared to with a GP.}
in situations such as the one considered in this article.

A number of other issues need to be investigated further:
the higher estimates of ECS when using the IID structure for some
of the ESMs, the higher uncertainty in the estimation of ECS when
using the IID structure for some of the ESMs, and the increased
uncertainty in the estimation of GP parameters as compared to that in
the estimation of IID or AR(1) parameters.

\bibliographystyle{ieeetr}
\bibliography{ci_abstract}

\begin{thebibliography}{1}

\bibitem{flato2013evaluation}
G.~Flato, J.~Marotzke, B.~Abiodun, P.~Braconnot, S.~C. Chou, W.~J. Collins,
  P.~Cox, F.~Driouech, S.~Emori, V.~Eyring, {\em et~al.}, ``Evaluation of
  climate models. in: Climate change 2013: The physical science basis.
  contribution of working group i to the fifth assessment report of the
  intergovernmental panel on climate change,'' {\em Climate Change 2013},
  vol.~5, pp.~741--866, 2013.

\bibitem{mitchell1990equilibrium}
J.~Mitchell, S.~Manabe, V.~Meleshko, and T.~Tokioka, ``Equilibrium climate
  change and its implications for the future,'' 1990.

\bibitem{roe2007climate}
G.~H. Roe and M.~B. Baker, ``Why is climate sensitivity so unpredictable?,''
  {\em Science}, vol.~318, no.~5850, pp.~629--632, 2007.

\bibitem{hoffert1980role}
M.~I. Hoffert, A.~J. Callegari, and C.-T. Hsieh, ``The role of deep sea heat
  storage in the secular response to climatic forcing,'' {\em Journal of
  Geophysical Research: Oceans}, vol.~85, no.~C11, pp.~6667--6679, 1980.

\bibitem{held2010probing}
I.~M. Held, M.~Winton, K.~Takahashi, T.~Delworth, F.~Zeng, and G.~K. Vallis,
  ``Probing the fast and slow components of global warming by returning
  abruptly to preindustrial forcing,'' {\em Journal of Climate}, vol.~23,
  no.~9, pp.~2418--2427, 2010.

\bibitem{geoffroy2013transient}
O.~Geoffroy, D.~Saint-Martin, D.~J. Olivi{\'e}, A.~Voldoire, G.~Bellon, and
  S.~Tyt{\'e}ca, ``Transient climate response in a two-layer energy-balance
  model. part i: Analytical solution and parameter calibration using cmip5
  aogcm experiments,'' {\em Journal of Climate}, vol.~26, no.~6,
  pp.~1841--1857, 2013.

\bibitem{hasselmann1976stochastic}
K.~Hasselmann, ``Stochastic climate models part i. theory,'' {\em Tellus},
  vol.~28, no.~6, pp.~473--485, 1976.

\end{thebibliography}

\end{document}